\documentclass{article}
\usepackage{spconf,amsmath,graphicx}
\usepackage{amsfonts,mathtools}
\usepackage{bm,bbm}
\usepackage{cite}
\usepackage{xcolor}
\usepackage{siunitx}
\usepackage{microtype}
\usepackage[subtle,title=normal,sections=normal,margins=normal,bibliography=normal]{savetrees}
\usepackage{pgfplots}
\usepackage{tikz}
\pgfplotsset{compat=newest}
\pgfplotsset{plot coordinates/math parser=false} 
\usepgfplotslibrary{patchplots}
\usepgfplotslibrary{colormaps}%
\newlength{\bibparskip}
\let\oldthebibliography\thebibliography
\renewcommand\thebibliography[1]{%
  \oldthebibliography{#1}%
  \setlength{\itemsep}{\bibparskip}%
}
\newcommand{\vectorADom}{\mathbf{a}\left(k,\theta_{\mathrm{d}}\right)}
\newcommand{\vectorGDom}{\mathbf{g}\left(k,\theta_{\mathrm{d}}\right)}
\newcommand{\vectorGhat}{\hat{\mathbf{g}}\left(k,l\right)}
\newcommand{\vectorXDom}{\mathbf{x}_{\mathrm{d}}\left(k,l\right)}
\newcommand{\vectorXDomDP}{\mathbf{x}_{\mathrm{d}}^{\mathrm{DP}}\left(k,l\right)}
\newcommand{\vectorXDomRev}{\mathbf{x}_{\mathrm{d}}^{\mathrm{R}}\left(k,l\right)}
\newcommand{\vectorXJ}{\mathbf{x}_{j}\left(k,l\right)}

\newcommand{\vectorXJDP}{\mathbf{x}_{j}^{\mathrm{DP}}\left(k,l\right)}
\newcommand{\vectorXJRev}{\mathbf{x}_{j}^{\mathrm{R}}\left(k,l\right)}
\newcommand{\vectorN}{\mathbf{n}\left(k,l\right)}
\newcommand{\vectorU}{\mathbf{u}\left(k,l\right)}

\newcommand{\vectorY}{\mathbf{y}\left(k,l\right)}
\newcommand{\vectorYherm}{\mathbf{y}^{H}\left(k,l\right)}
\newcommand{\phiY}{\boldsymbol{\Phi}_{\mathrm{y}}\left(k,l\right)}
\newcommand{\phiXDomDP}{\boldsymbol{\Phi}_{\mathrm{x}_{\mathrm{d}}}^{\rm{DP}}\left(k,l\right)}
\newcommand{\phiUDom}{\boldsymbol{\Phi}_{\mathrm{u}}\left(k,l\right)}
\newcommand{\hatphiY}{\hat{\boldsymbol{\Phi}}_{\mathrm{y}}\left(k,\hspace{1pt}l\right)}
\newcommand{\hatphiYPastFrame}{\hat{\boldsymbol{\Phi}}_{\mathrm{y}}\left(k,\hspace{1pt}l-1\right)}
\newcommand{\hatphiUDom}{\hat{\boldsymbol{\Phi}}_{\mathrm{u}}\left(k,\hspace{1pt}l\right)}
\newcommand{\hatphiUDomPastFrame}{\hat{\boldsymbol{\Phi}}_{\mathrm{u}}\left(k,\hspace{1pt}l-1\right)}
\newcommand{\indicatorFunctionKL}{\mathbbm{1}_{j}\left(k,l\right)}

\DeclarePairedDelimiter\curlyBrace\{\}
\newcommand{\myExpectation}[1]{\operatorname{\mathcal{E}}\curlyBrace*{#1}}
\DeclarePairedDelimiter\abs{\lvert}{\rvert}%
\DeclarePairedDelimiter\norm{\lVert}{\rVert}%
\title{Comparison of Frequency-Fusion Mechanisms For \\ Binaural Direction-of-Arrival Estimation for Multiple Speakers}
\name{Daniel Fejgin$^{\hspace{1pt}1}$, Elior Hadad$^{\hspace{1pt}2}$, Sharon Gannot$^{\hspace{1pt}2}$, Zbyn\v{e}k Koldovsk\'{y}$^{\hspace{1pt}3}$, Simon Doclo$^{\hspace{1pt}1}$\thanks{This work was funded by the Deutsche Forschungsgemeinschaft (DFG, German Research Foundation) under Germany's Excellence Strategy - EXC 2177/1 - Project ID 390895286 and Project ID 352015383 - SFB 1330 B2. The work was also supported by the Israeli Ministry of Science \& Technology. The work is also supported by the Department of the Navy, Office of Naval Research Global, through Project No.~N62909-23-1-2084.
}}
\address{$^1$University of Oldenburg and Cluster of Excellence Hearing4all, Oldenburg, Germany\\$^2$Bar-Ilan University, Ramat-Gan, Israel\\$^3$ Technical University of Liberec, Liberec, Czech Republic}
\begin{document}
\ninept
\maketitle
\begin{abstract}
To estimate the direction of arrival (DOA) of multiple speakers with methods that use prototype transfer functions, frequency-dependent spatial spectra (SPS) are usually constructed. To make the DOA estimation robust, SPS from different frequencies can be combined. According to how the SPS are combined, frequency fusion mechanisms are categorized into narrowband, broadband, or speaker-grouped, where the latter mechanism requires a speaker-wise grouping of frequencies. For a binaural hearing aid setup, in this paper we propose an interaural time difference (ITD)-based speaker-grouped frequency fusion mechanism. By exploiting the DOA dependence of ITDs, frequencies can be grouped according to a common ITD and be used for DOA estimation of the respective speaker. We apply the proposed ITD-based speaker-grouped frequency fusion mechanism for different DOA estimation methods, namely the multiple signal classification, steered response power and a recently published method based on relative transfer function (RTF) vectors. In our experiments, we compare DOA estimation with different fusion mechanisms. For all considered DOA estimation methods, the proposed ITD-based speaker-grouped frequency fusion mechanism results in a higher DOA estimation accuracy compared with the narrowband and broadband fusion mechanisms.
\end{abstract}
\section{Introduction}
\label{sec:intro}
In multi-microphone speech communication applications like hearing aids, estimation of the direction of arrival (DOA) of multiple speakers in noisy and reverberant environments is of crucial importance \cite{Doclo2015}. Due to their popularity, in this paper we specifically consider DOA estimation methods based on prototype transfer function matching. Examples for such methods include the well-established subspace-based multiple signal classification (\mbox{MUSIC}) \cite{Schmidt19886} or power-based steered response power (SRP) methods \cite{DiBiase2001}, which both use acoustic transfer function (ATF) vectors, or a recently published method based on matching of anechoic prototype re\-lative transfer function (RTF) vectors \cite{Fejgin2022}. In these methods, frequency-dependent spatial spectra (SPS), i.e., functions of candidate directions, are constructed with the location of peaks likely corresponding to the true speaker DOAs.

To make the DOA estimation of broadband speech sources robust, SPS from different frequencies can be combined. Depending on how the frequency-dependent SPS are combined, frequency fusion mechanisms are categorized into narrow\-band, broadband, or speaker-grouped \cite{Blandin2012,Thakallapalli2021}. With the narrowband fusion mechanism, i.e., no combination of frequency-dependent SPS, DOAs are first estimated directly for each frequency and subsequently clustered. From these clustered DOA estimates, refined estimates are obtained (e.g., \cite{Hadad2020}). With the broadband fusion mechanism, DOAs are estimated from a single SPS obtained from pooling frequency-dependent SPS. With the speaker-grouped fusion mechanism, DOAs are estimated from multiple SPS, each obtained from pooling frequency-dependent SPS. As each pooled SPS is associated with a single speaker solely, DOA estimation with the speaker-grouped fusion mechanism may overcome the tending over-emphasis of the dominant speaker and the lobe broadening of the spatial spectrum which may by introduced with the narrowband and broadband fusion mechanisms \cite{Blandin2012,Thakallapalli2021}. However, using the speaker-grouped fusion mechanism requires a speaker-wise grouping of frequencies.

In this paper, we propose an interaural time difference (ITD)-based speaker-grouped frequency fusion mechanism, which we apply to DOA estimation with the MUSIC, SRP and the recently published RTF-vector-matching-based methods. We exploit the DOA dependence of ITDs such that frequencies can be grouped according to a common ITD and, thus, be associated with a single speaker. We compare DOA estimation using the proposed ITD-based speaker-grouped fusion mechanism with the narrowband and broadband fusion mechanisms. To further assess the performance of DOA estimation using the proposed ITD-based speaker-grouped fusion mechanism, we also compare against DOA estimation with separated speakers, where the speakers are separated using successive applications of the independent vector extraction (IVE) algorithm \cite{Delfosse1995}. Experimental results with recorded data from the novel BRUDEX database \cite{Fejgin2023} for multiple reverberant and noisy acoustic scenarios with two static speakers and a binaural hearing aid setup demonstrate the efficiency of the proposed ITD-based speaker-grouped frequency fusion mechanism.

\section{Signal model and notation}
\label{sec:signalModel}
We consider a binaural hearing aid setup with $M$ microphones in a noisy and reverberant acoustic scenario with $J$ simul\-taneously active speakers $S_{1:J}$. We consider stationary DOAs $\theta_{1:J}$ (in the azimuthal plane) and $J$ assumed to be known. In the STFT domain, the $m$-th microphone signal can be written as\vskip-0.2cm
\begin{equation}
	Y_{m}\left(k,l\right) = \sum_{j=1}^{J}X_{m,j}\left(k,l\right) + N_{m}\left(k,l\right)\,,
	\label{eq:signalModel_micComponent}
\end{equation}
where $m\in\left\{1,\dots,M\right\}$, $k\in\left\{1,\dotsc,K\right\}$ and $l\in\left\{1,\dotsc,L\right\}$ denote the index of the microphone, frequency bin, and the frame, respectively, and $X_{m,j}\left(k,l\right)$ and $N_{m}\left(k,l\right)$ denote the $j$-th speech component and the noise component in the $m$-th microphone signal, respectively. Assuming disjoint speaker activity in the STFT domain \cite{Yilmaz2004}, each time-frequency (TF) bin is dominated by a single speaker across all microphone signals. \mbox{Stacking} all microphone signals into the $M$-dimensional vector $\vectorY =\left[Y_{1}\left(k,l\right),\,\dots,\,Y_{M}\left(k,l\right)\right]^{T}$ with $(\cdot)^{T}$ de\-noting the trans\-po\-sition operator, the vector of microphone signals can be approximated as $\vectorY \approx \vectorXDom + \vectorN$, with the vector $\vectorXDom$ denoting the dominant speech component and the vector $\vectorN$ denoting the noise component defined similarly as $\vectorY$.

We assume that each speech component $j$ can be split into a direct-path component $\vectorXJDP$ and a reverberation component $\vectorXJRev$, i.e., $\vectorXJ = \vectorXJDP + \vectorXJRev$. Assuming a multiplicative transfer function for $\vectorXJDP$ in the STFT domain \cite{Avargel2007}, the direct-path component $\vectorXDomDP$ can be written in terms of the direct-path acoustic transfer function (ATF) vector $\vectorADom$ (relating the microphone signals with the dominant speaker signal $S_{\mathrm{d}}\left(k,l\right)$ with DOA $\theta_{\mathrm{d}}$) or equivalently in terms of the direct-path relative transfer function (RTF) vector $\vectorGDom$ (relating the microphone signals with the reference microphone signal $X_{1,\mathrm{d}}\left(k,l\right)$) as\vskip-0.2cm\vskip1pt
\begin{equation}
	\vectorXDomDP = \vectorADom S_{\mathrm{d}}\left(k,l\right) = \vectorGDom X_{1,\mathrm{d}}\left(k,l\right)\,,
\end{equation}\vskip1pt\noindent
where the first microphone is considered as the reference microphone (without loss of generality). Condensing the noise and reverberation components into the undesired component $\vectorU = \vectorN + \vectorXDomRev$, where for conciseness the index $d$ of the dominant speaker was omitted in $\mathbf{u}$, the microphone signals can be written as $\vectorY = \vectorXDomDP + \vectorU$.

Assuming uncorrelated direct-path speech and undesired components, the covariance matrix $\phiY$ of the microphone signals can be approximated as\vskip-0.4cm
\begin{equation}
	\phiY = \myExpectation{\vectorY\vectorYherm} \approx \phiXDomDP+\phiUDom\,,
\end{equation}\vskip-0.2cm\noindent
where the covariance matrix $\phiXDomDP$ of the direct-path do\-minant speech component and the covariance matrix $\phiUDom$ of the undesired component are defined similarly as $\phiY$ and $(\cdot)^{H}$ and $\myExpectation{\cdot}$ denote the complex transposition and expectation operators, respectively.

\section{Frequency-dependent spatial spectra}
\label{sec:freqSPS}
In this section, we review the construction of frequency-dependent spatial spectra (SPS) for the subspace-based MUSIC \cite{Schmidt19886}, the power-based SRP \cite{DiBiase2001,Zohourian2018} and the RTF-vector-matching-based \cite{Fejgin2022} methods for binaural DOA estimation. For the construction of these frequency-dependent SPS a database of anechoic transfer functions for different candidate directions $\left\{\theta_{i}\right\}_{i=1}^{I}$ is used which accounts for head shadow effects. To obtain such a database, analytic diffraction models \cite{Duda1998} or measurements can be used.

For the MUSIC and SRP methods, the frequency-depen\-dent SPS are obtained from the estimated covariance matrix of the noisy microphone signals $\hatphiY$ and a database of anechoic head-related transfer functions $\bar{\mathbf{a}}\left(k,\theta_{i}\right)$. MUSIC is based on the ortho\-gonality between the noise subspace of $\phiY$ and the direct-path ATF vector $\vectorADom$, i.e.,\vskip-0.2cm
\begin{equation}
	\tilde{p}^{\mathrm{MUSIC}}\left(k,l,\theta_{i}\right) = \frac{1}{\norm{\hat{\boldsymbol{Q}}_{\mathrm{y},\mathrm{u}}^{H}\left(k,l\right)\bar{\mathbf{a}}\left(k,\theta_{i}\right)}_{2}}\,,
\end{equation}
where $\hat{\boldsymbol{Q}}_{\mathrm{y},\mathrm{u}}$ denotes the estimated noise subspace of $\hatphiY$. SRP is based on the maximization of the output power of a beamformer. \mbox{Accounting} for head shadow effects and the signal input power as in \cite{Zohourian2018}, the frequency-dependent SPS is given by
\begin{equation}
	\tilde{p}^{\mathrm{SRP}}\left(k,l,\theta_{i}\right) = \frac{\bar{\mathbf{a}}^{H}\left(k,\theta_{i}\right)\hatphiY \bar{\mathbf{a}}\left(k,\theta_{i}\right)}{\norm{\bar{\mathbf{a}}\left(k,\theta_{i}\right)}_{2}^{2}\norm{\vectorY}_{2}^{2}}\,.
\end{equation}

Similar to \cite{Salvati2014}, in our binaural ex\-periments with \mbox{MUSIC} and SRP we noted the im\-portance of normalizing the values of the frequency-dependent SPS to the range $\left[0,1\right]$. Hence, for binaural DOA estimation we consider the nor\-malized versions $p^{\mathrm{MUSIC}}\left(k,l,\theta_{i}\right)$ and $p^{\mathrm{SRP}}\left(k,l,\theta_{i}\right)$ instead of $\tilde{p}^{\mathrm{MUSIC}}\left(k,l,\theta_{i}\right)$ and $\tilde{p}^{\mathrm{SRP}}\left(k,l,\theta_{i}\right)$ in the following.
\begin{align}
	p^{\mathrm{MUSIC}}\left(k,l,\theta_{i}\right) &= \frac{\tilde{p}^{\mathrm{MUSIC}}\left(k,l,\theta_{i}\right)}{\underset{\theta_{j}}{\mathrm{max}}\hspace{0.1cm}\tilde{p}^{\mathrm{MUSIC}}\left(k,l,\theta_{j}\right)}\\
	p^{\mathrm{SRP}}\left(k,l,\theta_{i}\right) &= \frac{\tilde{p}^{\mathrm{SRP}}\left(k,l,\theta_{i}\right)}{\underset{\theta_{j}}{\mathrm{max}}\hspace{0.1cm}\tilde{p}^{\mathrm{SRP}}\left(k,l,\theta_{j}\right)}\,.
\end{align}\vskip-0.2cm

For the RTF-vector-matching-based method, the frequency-dependent SPS is obtained from the comparison between estimated RTF vector $\vectorGhat$ and a database of anechoic prototype RTF vectors $\bar{\mathbf{g}}\left(k,\theta_{i}\right)$ for different candidate directions $\theta_{i}$. Considering the \mbox{Hermitian} angle for the com\-parison of RTF vectors, the frequency-dependent SPS is obtained as\vskip-0.2cm
\begin{equation}
	p^{\mathrm{RTF}}\left(k,l,\theta_{i}\right) = - \mathrm{arccos}\left(\frac{\abs{\bar{\mathbf{g}}^{H}\left(k,\theta_{i}\right)\vectorGhat}}{\norm{\bar{\mathbf{g}}\left(k,\theta_{i}\right)}_{2}\norm{\vectorGhat}_{2}}\right)
\end{equation}
To estimate the DOAs $\hat{\theta}_{1:J}\left(l\right)$, assuming that the number of speaker $J$ is known, we consider combining the frequency-dependent SPS $p\left(k,l,\theta_{i}\right)$ using the frequency-fusion mechanisms which are discussed in Section \ref{sec:freqFusion}.

\section{Combination of frequency-dependent spatial spectra for binaural DOA estimation}
\label{sec:freqFusion}
In this section, we first review frequency subset selection criteria and principles of narrow\-band, broadband, and speaker-grouped frequency-fusion mechanisms for DOA estimation (Sections \ref{subsec:freqFusion__subsetSelection} and \ref{subsec:freqFusion__principles}). In Section \ref{subsec:freqFusion__ITD}, we propose for the speaker-grouped frequency-fusion mechanisms a frequency grouping that is based on estimated interaural time differences (ITDs).

\subsection{Frequency subset selection}
\label{subsec:freqFusion__subsetSelection}
Assuming that DOAs can only be reliably estimated at frequencies where one speaker dominates over all other speakers, noise and reverberation, we restrict DOA estimation to the subset $k\in\mathcal{K}\left(l\right)$ which due to time-varying microphone recordings is also time-varying. In the context of DOA estimation, criteria based on, e.g., speaker dominance \cite{Madmoni2019}, signal-to-noise ratio (SNR) \cite{Tho2014,Hadad2020}, speech onsets \cite{Tho2014}, or coherence-based quantities such as the coherent-to-diffuse ratio (CDR) \cite{Taseska2017,Fejgin2022} have been proposed for frequency subset selection. In this paper, we consider the so-called binaural effective-coherence-based coherent-to-diffuse ratio (CDR) criterion
\begin{equation}
	\mathcal{K}\left(l\right) = \left\{k: \widehat{\mathrm{CDR}}\left(k,l\right)\geq\mathrm{CDR}_{\mathrm{thresh}}\right\}\,,\label{eq:CDRcriterion}
\end{equation}\vskip-0.1cm\noindent
which is based on a binaural coherence model that accounts for head-shadowing effects \cite{Lindevald1986} and the average coherence between signals from all possible microphone pairs between the left and the right hearing aid. The detailed computation of $\widehat{\mathrm{CDR}}\left(k,l\right)$ is described in \cite{Fejgin2022}.

\subsection{Frequency fusion mechanisms for DOA estimation}
\label{subsec:freqFusion__principles}
With the narrowband fusion mechanism, i.e., no combination of frequency-dependent SPS, the frequency-dependent SPS $p\left(k,l,\theta_{i}\right)$ are maximized directly per frequency, i.e.,
\begin{equation}
    \hat{\theta}\left(k,l\right) = \underset{\theta_{i}}{\mathrm{argmax}}\hspace{0.1cm}p\left(k,l,\theta_{i}\right)\,,
\end{equation} 
where due to the assumed disjoint speaker activity \cite{Yilmaz2004} only a single DOA is estimated per frequency. Subsequently, the DOA estimates of the frequency subset $k\in\mathcal{K}\left(l\right)$ are combined using, e.g., a histogram operation $\mathcal{H}$ \cite{Hadad2020} which counts how many DOA estimates correspond to each candidate direction $\theta_{i}$. The speaker DOAs are estimated as the location of peaks of this histogram, i.e.,
\begin{equation}
	\hat{\theta}_{1:J}^{\mathrm{narrow}}\left(l\right) = \underset{\theta_{i}}{\mathrm{findPeaks}}\hspace{0.1cm}\mathcal{H}\left\{\hat{\theta}\left(k,l\right)\right\}\,, \quad k\in\mathcal{K}\left(l\right)\,.
\end{equation}

With the broadband frequency-fusion mecha\-nism, the frequency-dependent SPS $p\left(k,l,\theta_{i}\right)$ are combined using a pooling o\-pe\-ration, e.g., a summation \cite{Fejgin2022}. The speaker DOAs are estimated as the location of peaks of this pooled SPS, i.e.,
\begin{equation}
	\hat{\theta}_{1:J}^{\mathrm{broad}}\left(l\right) = \underset{\theta_{i}}{\mathrm{findPeaks}}\hspace{0.1cm}\sum_{k}\hspace{2pt}p\left(k,l,\theta_{i}\right)\,, \quad k\in\mathcal{K}\left(l\right)\,.
\end{equation}

With the speaker-grouped frequency-fusion mechanism, in\-di\-vi\-dual frequencies are associated with a single speaker solely, allowing to construct for each individual speaker its own SPS. Let the indicator function $\indicatorFunctionKL$ denote the \mbox{association} between TF bin $\left(k,l\right)$ and the $j$-th speaker, i.e.,
\begin{equation}
	\indicatorFunctionKL = \begin{cases}
		1 & \text{TF bin }\left(k,l\right)\text{ associated with speaker $j$}\\
		0 & \text{else}\,.
	\end{cases}
	\label{eq:TFassociation_general}
\end{equation}
Only frequency-dependent SPS $p\left(k,l,\theta_{i}\right)$ that are associated with the $j$-th speaker are combined using a pooling operation. The $j$-th speaker DOA is estimated as the location of the maximum of the $j$-th pooled SPS, i.e., for $j=1,...,J$
\begin{equation}
	\hat{\theta}_{j}^{\mathrm{group}}\left(l\right) = \underset{\theta_{i}}{\mathrm{argmax}}\hspace{0.1cm}\sum_{k}\hspace{2pt}\indicatorFunctionKL\hspace{2pt}p\left(k,l,\theta_{i}\right)\,, \quad k\in\mathcal{K}\left(l\right)\,.
\end{equation}

\subsection{ITD-based speaker-grouped frequency grouping}
\label{subsec:freqFusion__ITD}
In order to compute the association $\indicatorFunctionKL$ in \eqref{eq:TFassociation_general}, we want to exploit the spatial information provided by mul\-tiple microphone signals of the binaural hearing aid setup.
Since spatially disjoint \mbox{speakers} can be discriminated by their interaural time differences (ITDs), we propose to combine the method for the estimation of ITDs from \cite{ElChami2009} with the estimation of relative transfer functions (RTFs) and subsequently group frequencies accor\-ding to common ITDs. It should be noted that ITDs are a popular feature for frequency grouping and binaural DOA estimation \cite{Rickard2000,Mandel2010,May2012}. However, exploiting ITD-based frequency grouping for DOA estimation using the MUSIC, SRP and RTF-vector-matching-based methods have not been con\-sidered.

The main assumption of our ITD-based frequency grouping is that the interaural phase difference (IPD) between a pair of contra-lateral microphone signals changes linearly with frequency, with the slope of this linear relation corresponding to the ITD which depends on the speaker DOA. For the estimation of ITDs we consider the grid-search approach with candidate ITDs $\tau_{n}$ as in \cite{ElChami2009}, where we compare the phase $\mathrm{arg}\left\{\hat{G}\left(k,l\right)\right\}$ of the estimated RTF between a pair of contra-lateral microphone signals against a set of phase differences with candidate ITDs $\tau_{n}$, i.e., $\mathrm{arg}\left\{\mathrm{e}^{\mathrm{i}\,\omega_{k}\tau_{n}}\right\}$, with $\omega_{k}$ denoting the discrete angular frequency at frequency bin with index $k$. Accounting for the phase wrapping due to the phase extraction operator $\mathrm{arg}\left\{\cdot\right\}$ and considering the non-linear trans\-formation $\rho\left(x\right) = \mathrm{e}^{\beta\cos\left(x\right)} / \mathrm{e}^{\beta}$ with hyper\-parameter $\beta$, a score function $\Psi\left(k,l,\tau_{n}\right)$ in $\tau_{n}$ is constructed, i.e.,
\begin{equation}
	\Psi\left(k,l,\tau_{n}\right) = \rho\left(\mathrm{arg}\left\{\hat{G}\left(k,l\right)\right\} - \mathrm{arg}\left\{\mathrm{e}^{\mathrm{i}\,\omega_{k}\tau_{n}}\right\}\right)\,.
\end{equation}
Assuming that the number of speakers $J$ is known, the scores are frequency-averaged and the $J$ ITDs, each associated with one of the $J$ speakers, are estimated as the location of peaks of the average score function as\vskip-0.2cm
\begin{align}
    \hat{\tau}_{1:J}\left(l\right) &= \underset{\tau_{n}}{\mathrm{findPeaks}}\hspace{2pt}\sum_{k\in\mathcal{K}\left(l\right)}\hspace{2pt}\Psi\left(k,l,\tau_{n}\right)
\end{align}\vskip-0.1cm

For the estimation of the indicator function $\indicatorFunctionKL$ we propose a TF-wise maximization of the similarity scores $\Psi\left(k,l,\hat{\tau}_{j}\right)$ evaluated using each estimated ITD $\hat{\tau}_{1:J}\left(l\right)$, i.e.,\vskip-0.1cm
\begin{equation}
	\boxed{\indicatorFunctionKL = \begin{cases}
			1 & \text{if } j= \underset{j^{\prime}\in\left\{1,...,J\right\}}{\mathrm{argmax}}\,\Psi\left(k,l,\hat{\tau}_{j^{\prime}}\left(l\right)\right)\\
			0 & \text{else}
	\end{cases}}
	\label{eq:TFassociation_specific}
\end{equation}\vskip-0.1cm

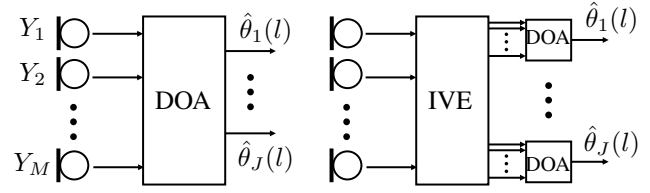
\begin{figure}[t]
	\centering
		\begin{tikzpicture}
			\node [align=center,scale=1]
			{\includegraphics[width=0.45\columnwidth,trim={6cm 5cm 6cm 5cm},clip]{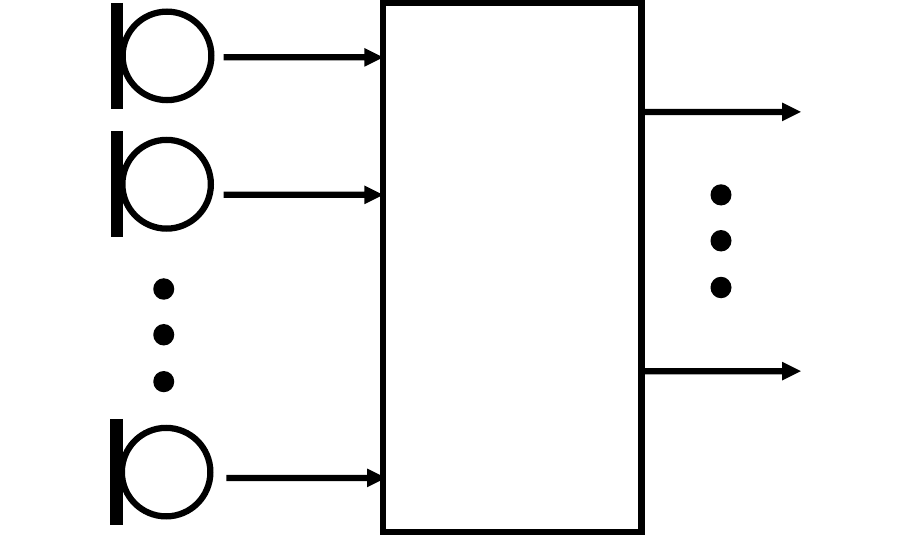}\hskip0.5em\includegraphics[width=0.45\columnwidth,trim={6cm 5cm 6cm 5cm},clip]{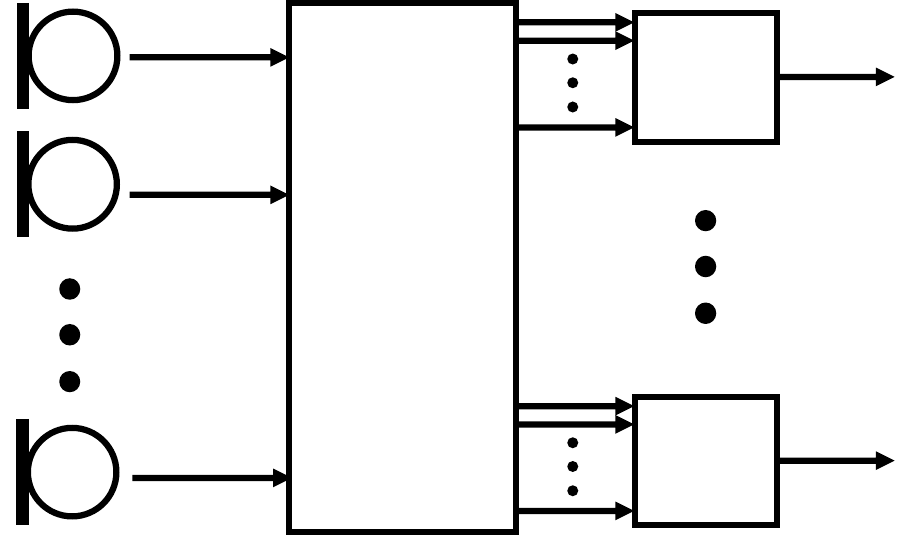}};
			\node [xshift=-3.8cm,yshift=0.95cm,font = \fontsize{10}{1}\selectfont,] {$Y_{1}$};
			\node [xshift=-3.8cm,yshift=0.35cm,font = \fontsize{10}{1}\selectfont,] {$Y_{2}$};
			\node [xshift=-3.8cm,yshift=-0.85cm,font = \fontsize{10}{1}\selectfont,] {$Y_{M}$};
			\node [xshift=-1.8cm,yshift=0cm,font = \fontsize{10}{1}\selectfont,] {DOA};
			\node [xshift=-0.7cm,yshift=0.95cm,font = \fontsize{10}{1}\selectfont,] {$\hat{\theta}_{1}\left(l\right)$};
			\node [xshift=-0.7cm,yshift=-0.75cm,font = \fontsize{10}{1}\selectfont,] {$\hat{\theta}_{J}\left(l\right)$};
			\node [xshift=1.75cm,yshift=0cm,font = \fontsize{10}{1}\selectfont,] {IVE};
			\node [xshift=3.05cm,yshift=0.84cm,font = \fontsize{8}{1}\selectfont,] {DOA};
			\node [xshift=3.05cm,yshift=-0.84cm,font = \fontsize{8}{1}\selectfont,] {DOA};
			\node [xshift=3.9cm,yshift=1.1cm,font = \fontsize{10}{1}\selectfont,] {$\hat{\theta}_{1}\left(l\right)$};
			\node [xshift=3.9cm,yshift=-0.55cm,font = \fontsize{10}{1}\selectfont,] {$\hat{\theta}_{J}\left(l\right)$};
	\end{tikzpicture}
    \vspace{-3mm}
	\caption{Comparison of DOA estimation without (left) and with (right) application of a blind speaker-separation algorithm.\label{fig:blockDiagram}}
    \vskip-3.4mm
\end{figure}

\section{Speaker-separation-based DOA estimation}
\label{sec:IVEbasedDOA}
To further assess the performance of DOA estimation using the proposed ITD-based speaker-grouped frequency fusion mechanism, in our ex\-pe\-riments in Section \ref{sec:experiments} we also consider DOA estimation with blindly separated speakers. It should be noted that DOA estimation based on blind speaker separation algorithms has previously been investigated in, e.g., \cite{Lombard2011}.

For the speaker se\-pa\-ration, we consider successive applications of the independent vector extraction (IVE) algorithm, which assumes independent and known amount of speakers and exploits inter-frequency dependencies \cite{Delfosse1995}. The speaker separation is performed offline using the entire mixture. For further details regarding the speaker sepation algorithm, we refer the readers to \cite{Delfosse1995} and \cite{Koldovsky2021} and references therein. For each separated speaker the DOA is estimated using a method from Section \ref{sec:freqSPS} using the broadband fusion mechanism described in Section \ref{subsec:freqFusion__principles}. In Fig.~\ref{fig:blockDiagram} the concept of DOA estimation with and without separated speakers is visualized.

\section{Experimental results}
\label{sec:experiments}
For static two-speaker acoustic scenarios in multiple rever\-berant environments with diffuse-like babble noise, in this section we compare the DOA estimation performance for the MUSIC, SRP, and the RTF-vector-matching-based DOA estimation methods when applied with the narrow\-band, broadband, and the proposed ITD-based speaker-grouped frequency fusion mechanisms. We also consider DOA estimation with se\-parated speakers using successive applications of the IVE algorithm. In Section \ref{subsec:experiments_setup}, we describe the experimental setup and implementations details. In Section \ref{subsec:experiments_results} we present and discuss the results in terms of DOA estimation accuracy.

\subsection{Experimental setup and implementation details}
\label{subsec:experiments_setup}
For the experiments, we consider separate recordings of reverberant speech and diffuse-like babble noise using hearing aid microphones from the BRUDEX database \cite{Fejgin2023}. We consider three rever\-beration environments ('low', 'medium', and 'high'), corresponding to median rever\-beration times $\mathrm{T}_{60}\approx\left[240,485,1170\right]\,\mathrm{ms}$. Excluding co-located speakers, we consider 132 possible DOA combinations in the range $\left[-150:30:180\right]\,^{\circ}$ of a female and a male speaker. We cut the speech signals to a duration of approxi\-mately \qty{5}{\second} and use an oracle energy-based voice activity detector to obtain constantly active speech signals over the considered duration. We consider equal ave\-rage broadband speech power across all microphones for both speakers. The noise component is scaled according to SNRs in the range $\left[-5:5:20\right]\,\mathrm{dB}$, where the SNR is set as the ratio of the average broadband speech power of one speaker across all microphones to the average broadband noise power across all microphones. 

All microphone signals are downsampled to \qty{16}{\kilo\hertz} and processed within an STFT framework with \qty{32}{\milli\second} square-root Hann windows with \qty{50}{\percent} overlap. We estimate the cova\-riance matrices $\hatphiY$ of the noisy microphone signals and $\hatphiUDom$ of the undesired component for each TF bin using a first-order recursion during speech-and-noise periods and noise-only periods, respectively, as
\begin{align}
	\hatphiY &= \alpha_{\mathrm{y}}\hatphiYPastFrame + \left(1-\alpha_{\mathrm{y}}\right) \mathbf{y}\left(k,\hspace{1pt}l\right)\mathbf{y}^{H}\left(k,\hspace{1pt}l\right)\label{eq:SOS_RyUpdate}\\
	\hatphiUDom &= \alpha_{\mathrm{u}}\hatphiUDomPastFrame + \left(1-\alpha_{\mathrm{u}}\right) \mathbf{y}\left(k,\hspace{1pt}l\right)\mathbf{y}^{H}\left(k,\hspace{1pt}l\right)\label{eq:SOS_RnUpdate}\,,
\end{align}
with smoothing factors $\alpha_{\mathrm{y}}$ and $\alpha_{\mathrm{u}}$ corresponding to time constants of \qty{250}{\milli\second} and \qty{500}{\milli\second}, respectively. With a similar recursion the covariance matrices of the separated speech components $\hat{\boldsymbol{\Phi}}_{\hat{\mathrm{x}},j}\left(k,\hspace{1pt}l\right)$ and noise components $\hat{\boldsymbol{\Phi}}_{\mathrm{u},j}\left(k,\hspace{1pt}l\right)$ are estimated. To discriminate speech-and-noise periods from noise-only periods, a speech presence pro\-ba\-bility \cite{Gerkmann2012} is estimated in all microphones, averaged and thresholded.

We estimate the RTF vector using the state-of-the-art covariance whitening method \cite{Markovich2009,MarkovichGolan2018}. Only the entry of the RTF vector that relates the front hearing aid microphone signals with each other is considered for the ITD-based speaker-grouped frequency-fusion mechanism. Psycho-acoustically motivated, we consider candidate ITDs in the range $\left[-900:1:900\right]\,\si{\micro\second}$, and we set the hyperparameter $\beta=5$. The IVE algorithm is implemented accor\-ding to \cite{Koldovsky2021} with at most 100 iterations.

For the computation of the frequency-dependent SPS, we consider measured anechoic binaural room impulse responses with an angular resolution of \qty{5}{\degree} in the range $\left[-180:5:175\right]^{\circ}$ \cite{Kayser2009} from which the sets of transfer functions $\left\{\bar{\mathbf{a}}\left(k,\theta_{i}\right)\right\}_{i=1}^{I}$ and $\left\{\bar{\mathbf{g}}\left(k,\theta_{i}\right)\right\}_{i=1}^{I}$  are obtained for $I=72$ candidate directions.

We assess the DOA estimation performance using the accuracy which we define as:
\begin{equation}
	\mathrm{ACC} = \frac{1}{L}\sum_{l=1}^{L}\mathrm{ACC}\left(l\right); \quad \mathrm{ACC}\left(l\right) = j_{\mathrm{correct}}\left(l\right)/J\,,
\end{equation}
where $j_{\mathrm{correct}}\left(l\right)$ denotes the number of speakers in the $l$-th frame for which the DOA is estimated within $\pm 5^{\circ}$ correctly. For each DOA estimation procedure, i.e., the combination of DOA estimation method, frequency fusion mechanism, and (non-) application of speaker separation algorithm, we compute the median of accuracies taken over all acoustic scenes which are specified by the DOA  pair, the SNR, and the reverberation condition.
 
\subsection{Results}
\label{subsec:experiments_results}
\begin{figure}[t]%
	\centering
%
%
\begin{tikzpicture}[scale=0.195]

\begin{axis}[%
width=15.5in,
height=8.175in,
at={(2.6in,1.103in)},
scale only axis,
bar shift auto,
xmin=0.2,
xmax=4.2,
xtick={{0.65},{1.65},{2.65},{3.65}},
xticklabels={{narrowband},{broadband},{ITD-based},{IVE-based}},
xlabel style={font=\color{black}, font = \fontsize{45}{1}\selectfont},
xticklabel style={font=\color{black}, font = \fontsize{45}{1}\selectfont, rotate=0,align=center,yshift=-0.25cm},
ymin=50,
ymax=100,
ytick={{0},{5},{10},{15},{20},{25},{30},{35},{40},{45},{50},{55},{60},{65},{70},{75},{80},{85},{90},{95},{100}},
yticklabels={{0},{},{10},{},{20},{},{30},{},{40},{},{50},{},{60},{},{70},{},{80},{},{90},{},{100}},
ylabel style={font=\color{black}, font = \fontsize{45}{1}\selectfont},
yticklabel style={font=\color{black}, font = \fontsize{45}{1}\selectfont},
ylabel={Accuracy [\%]},
axis background/.style={fill=white},
grid=both,
grid style={black,line width=0.8pt},
axis line style = {line width=2pt},
legend image post style={scale=3},
legend columns=-1,
legend entries={Long plot title, B, C},
legend style={/tikz/every even column/.append style={column sep=2cm}},
legend style={at={(0.01,0.99)}, anchor=north west, legend cell align=left, align=left, draw=none, font = \fontsize{40}{1}\selectfont}
]
\addplot[ybar, bar width=0.222, fill=blue, draw=black, area legend] table[row sep=crcr] {%
	1	67.5101091298841\\
	2	72.8253191074734\\
	3	79.7665156387021\\
	4	85.9245590957809\\
};
\addlegendentry{MUSIC}
\addplot[ybar, bar width=0.222, fill=black, draw=black, area legend] table[row sep=crcr] {%
	1	76.1132222547014\\
	2	66.6587498782033\\
	3	75.1504189808048\\
	4	76.2551154633148\\
};
\addlegendentry{SRP}
\addplot[ybar, bar width=0.222, fill=red, draw=black, area legend] table[row sep=crcr] {%
	1	78.8621748026893\\
	2	68.6635243106304\\
	3	79.2074685764397\\
	4	78.4815599727175\\
};
\addlegendentry{RTF-based}
%
\addplot[ybar, bar width=0.111, fill=blue, draw=white, area legend, line width = 4pt] table[row sep=crcr] {%
	0.368	61.8770096463023\\
	1.368	60.3941342687323\\
	2.368	70.9703546721232\\
	3.368	72.9044869921076\\
};
\addplot[ybar, bar width=0.111, fill=black, draw=white, area legend, line width = 4pt] table[row sep=crcr] {%
	0.479	72.2955032641528\\
	1.479	61.0895936860567\\
	2.479	72.5330069180552\\
	3.479	73.7132173828315\\
};
\addplot[ybar, bar width=0.111, fill=red, draw=white, area legend, line width = 4pt] table[row sep=crcr] {%
	0.5890	76.5949283835136\\
	1.5890	62.2813748416642\\
	2.5890	78.4779060703498\\
	3.5890	74.8836841079606\\
};
\node[above, align=center,font = \fontsize{32.5}{1}\selectfont,color=blue]
at (axis cs:0.35,92) {-3\,dB};
\node[above, align=center,font = \fontsize{32.5}{1}\selectfont,color=black]
at (axis cs:0.65,92) {-5\,dB};
\node[above, align=center,font = \fontsize{32.5}{1}\selectfont,color=red]
at (axis cs:0.95,92) {-5\,dB};
\node[above, align=center,font = \fontsize{32.5}{1}\selectfont,color=blue]
at (axis cs:1.35,92) {-2\,dB};
\node[above, align=center,font = \fontsize{32.5}{1}\selectfont,color=black]
at (axis cs:1.65,92) {-1\,dB};
\node[above, align=center,font = \fontsize{32.5}{1}\selectfont,color=red]
at (axis cs:1.95,92) {-1\,dB};
\node[above, align=center,font = \fontsize{32.5}{1}\selectfont,color=blue]
at (axis cs:2.35,92) {-3\,dB};
\node[above, align=center,font = \fontsize{32.5}{1}\selectfont,color=black]
at (axis cs:2.65,92) {-4\,dB};
\node[above, align=center,font = \fontsize{32.5}{1}\selectfont,color=red]
at (axis cs:2.95,92) {-5\,dB};
\node[above, align=center,font = \fontsize{32.5}{1}\selectfont,color=blue]
at (axis cs:3.35,92) {4\,dB};
\node[above, align=center,font = \fontsize{32.5}{1}\selectfont,color=black]
at (axis cs:3.65,92) {5\,dB};
\node[above, align=center,font = \fontsize{32.5}{1}\selectfont,color=red]
at (axis cs:3.95,92) {4\,dB};
\end{axis}

\end{tikzpicture}%
%
	\vspace{-3mm}
	\caption{Median DOA estimation accuracy for the investigated DOA estimation procedures. The white bars denote the median accuracy without frequency subset selection and the number above each bar indicates the CDR threshold value leading to the highest median accuracy.\label{fig:results}}
	\vskip-3.4mm
\end{figure}
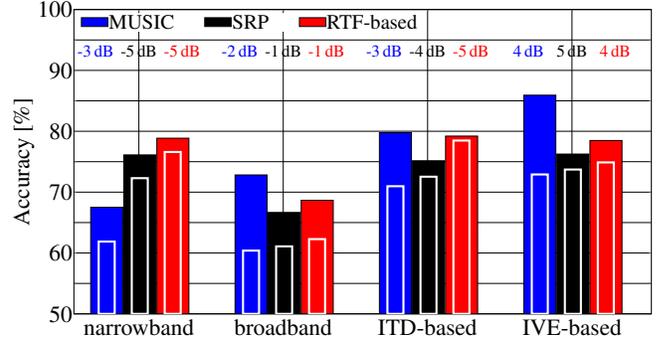
In Fig.~\ref{fig:results}, we compare the median DOA estimation accuracy for the considered DOA estimation procedures. For each procedure we also compare the median DOA estimation accuracy when selecting only a subset of frequencies (corresponding to the threshold value $\mathrm{CDR}_{\mathrm{thresh}}$ \mbox{leading} to the highest median accuracy, where the respective threshold value is indicated above each bar) against when using the whole frequency set (corres\-ponding to $\mathrm{CDR}_{\mathrm{thresh}} = \qty[parse-numbers = false]{-\infty}{\decibel}$, shown as narrow white bars).

First, it can be observed that similar as in \cite{Fejgin2022} for all considered DOA estimation procedures the frequency subset selection improves DOA estimation, with the largest impact with the MUSIC method. Se\-cond, comparing the procedures from the narrowband and broadband frequency fusion methods, variations in the estimation accuracy up to \qty{10}{\percent} for the same DOA estimation method can be observed. Third, com\-pa\-ring the proposed ITD-based speaker-grouped frequency fusion mechanism with the narrowband and broadband frequency fusion mechanism, for almost all DOA estimation procedures an increase in localization accuracy can be observed. Comparing the speaker-grouped frequency fusion mechanism against the broadband fusion mechanism, the accuracy can be increased by at least \qty{8}{\percent} for all DOA estimation methods. Comparing the speaker-grouped frequency fusion mechanism against the narrowband fusion mechanism, for the MUSIC method the accuracy can be increased by about \qty{12}{\percent}, for the RTF-vector-matching-based method a minor improvement can be achieved, and for the SRP method there is no improvement, indicating that both speakers can already be detected in the SPS of the latter methods. Fourth, comparing the procedures from the proposed online ITD-based speaker-grouped frequency fusion mechanism with the procedures involving application of the batch offline IVE algorithm, one sees comparable results for the SRP and RTF-vector-matching-based me\-thods and an improvement of \qty{6}{\percent} for the MUSIC method. In summary, our experiments highlight the effectiveness of the proposed ITD-based speaker-grouped frequency fusion mechanism for DOA estimation.

\section{Conclusion}
\label{sec:conclusion}
In this paper, we compared how the combination of frequency-dependent spatial spectra affect DOA estimation for a binaural hearing aid setup. More in particular, we proposed a speaker-grouped frequency fusion mechanism based on ITDs and compared it against a narrowband and broadband frequency fusion mechanism using the \mbox{MUSIC}, SRP, and RTF-vector-based prototype matching DOA estimation methods. To further assess DOA estimation using the proposed ITD-based speaker-grouped frequency fusion mechanism, we also considered DOA estimation with blindly separated speakers. Our experimental results based on noisy and reverberant re\-cor\-dings from the BRUDEX-database demonstrate the effectiveness of the proposed ITD-based speaker-grouped frequency fusion mechanism.
%
%
%
\bibliographystyle{IEEEbib}
\bibliography{refs}
\end{document}